\documentclass[12pt]{article}
\usepackage{epsfig}
\usepackage{amsfonts}
\usepackage{amsmath}
\usepackage{amssymb}
\usepackage{cite}
\usepackage{rotating}
\begin{document}
\sffamily
\title{
Nuclear fusion induced by X-rays in a crystal
}
\author{
\fbox{V.B. Belyaev${}^1$}
\footnote{Passed away in March 2015}
\ , M.B. Miller${}^2$, J. Otto${}^3$, S.A.
Rakityansky${}^3$
\footnote{e-mail: rakitsa@up.ac.za}
\\[3mm]
\parbox{11cm}{%
${}^1$ {\small Joint Institute for Nuclear Research, Dubna, Russia}\\
${}^2$ {\small Institute for Physical and Technical Problems, Dubna,
                      Russia}\\
${}^3$ {\small Dept. of Physics, University of Pretoria, Pretoria,
               South Africa}
}}
\maketitle
\begin{abstract}
\noindent
The nuclei that constitute a crystalline lattice, oscillate relative to
each other with a very low energy that is not sufficient to penetrate through
the Coulomb barriers separating them. An additional energy, which is needed to
tunnel through the barrier and fuse, can be supplied by external electromagnetic
waves (X-rays or the synchrotron radiation). Exposing to the X-rays the solid
compound LiD (lithium-deuteride) for the duration of 111 hours, we have
detected 88 events of the nuclear fusion
$d+{}^6\mathrm{Li}\to{}^8\mathrm{Be}^*$. Our theoretical estimate agrees with
what we observed. One of possible applications of the phenomenon we found,
could be  the measurements of the rates of various nuclear reactions (not
necessarily fusion) at extremely low energies inaccessible in accelerator
experiments.
\end{abstract}

\section{Introduction}
Fusion of two atomic nuclei is possible if they approach each other to a short
distance ($\sim~\!\!\!10^{-13}$\,cm). To come that close, they need to go
through a Coulomb barrier of the height of few MeV. The
penetration probability for such a barrier at room temperature (energy of
relative motion $\sim~\!\!\!10$\,meV) is practically zero
($\sim~\!\!\!10^{-2600}$ \cite{Balian}), but this probability rapidly grows when
the kinetic energy of the nuclei increases. For example, for the $dd$
system at the energy of 30\,keV the penetration probability
becomes $\sim~\!\!\!10^{-3}$ \cite{Balian}. Therefore, the obvious way to
fuse the nuclei is to raise the temperature of their mixture. In this way the so
called thermo-nuclear reactions happen in the stellar bodies, in nuclear
weapons, and in the TOKAMAK\cite{tokamak}.\\

Alternatively, the nuclei (which we want to fuse) can be put in a bound system
such as a molecule, where they sit next to the barrier relatively
long. In such a case, even at zero relative energy, the nuclei are separated by
a thinner barrier. For example, the two deuterons sitting at rest near each
other at a distance of $1\mathrm{\AA}$, have to overcome the same barrier as two
free moving deuterons with relative energy of $14$\,eV (which is equivalent to
the temperature of $\sim 1.6\times10^5\,{}^\circ\mathrm{K}$).\\

This, however, does not help much. For a deuterium molecule $\mathrm{D}_2$ in
its ground state the penetration probability is still too low, namely,
$\sim~\!\!\!10^{-82}$ \cite{Balian} and the fusion rate is
$\sim~\!\!\!10^{-62}\,\mathrm{s^{-1}}$ \cite{Picker}. These numbers can be
significantly increased if we make the size of the molecule smaller.
This can be achieved if
one electron in the molecule is replaced with the muon, which is
approximately 200 times heavier. As a result, the nuclei find themselves at a
distance that is 200 times smaller \cite{Jackson, Rafelski}. For the deuterons,
the thickness of the barrier becomes the same as at the collision energy almost
equal to 3\,keV (this corresponds to the temperature of
$\sim30\times10^6\,\mathrm{{}^\circ K}$).\\

This is already a significant gain.
The penetration probability for the muonic molecule $dd\mu$ is
$\sim~\!\!\!10^{-6}$ \cite{Balian, JINR}. With such a probability, one muon can
help to fuse, i.e. can catalyze the fusion of many nuclear pairs before it
decays (muon lifetime is $\sim~\!\!\!2\times10^{-6}$\,s).
The muon-catalyzed fusion has been observed and well studied both
experimentally and theoretically
(see, for example, Refs.\cite{Jones, JINR}), but turned out to be
inefficient as a new source for the energy production.\\

In the present paper, we suggest and experimentally explore yet another possible
approach to fusion of light nuclei. The idea is to make a crystal out of the
atoms whose nuclei we want to fuse. In this crystal, the nuclei sit next to each
other at an atomic distance and oscillate around the equilibrium positions. A
crystal is just a huge molecule and of course
the probability of spontaneous fusion of neighbouring nuclei is negligible, the
same as in the ordinary molecules. An experiment
aimed to observe the spontaneous fusion in the lithium deuteride crystal put an
upper bound on the fusion rate per nuclear pair as
$\sim10^{-48}\,\mathrm{s}^{-1}$ \cite{BelyaevFBS2006}.\\

However, we can try to shake the crystalline lattice by an external force
(electromagnetic
wave, for example). The nuclei swayed around their equilibrium positions,
acquire kinetic energy relative to each other. As a result, the original
Boltzmann distribution of the nuclei over the oscillation levels is changing
and the higher levels are populated. From these higher energy levels
the nuclei can tunnel through the Coulomb barrier and fuse. This is what
we observed when irradiated  the LiD crystal with the X-rays. The rate of
such a fusion turned out to be very low (one event every half an hour),
but still easily measurable.

\section{Nuclear subsystem}
\subsection{Fusion channels}
\label{sec.FusionChannels}
\begin{figure}[ht!]
\centerline{\epsfig{file=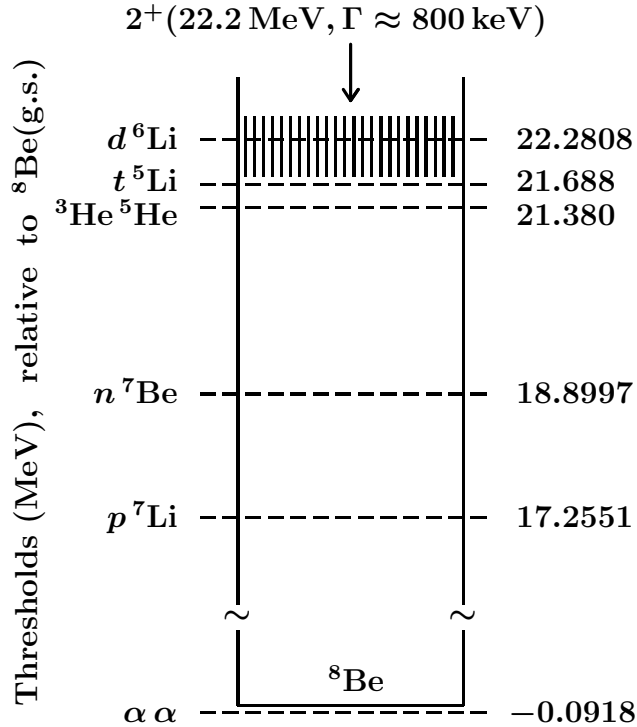}}
\caption{\sf
Threshold energies (in MeV) for various two-cluster arrangements of the nucleons
 constituting ${}^8\mathrm{Be}$ nucleus. The energies are shown relative to the
ground state of ${}^8\mathrm{Be}$. A wide meta-stable state ($2^+$) of this
nucleus  is shown around the $d\,{}^6\mathrm{Li}$ threshold.
}
\label{fig.8Be_spectrum}
\end{figure}

The nuclei we want to fuse, are the isotopes of hydrogen and lithium, namely,
${}^2\mathrm{H}$ and ${}^6\mathrm{Li}$. Their compound nucleus,
${}^8\mathrm{Be}$, has no stable states. The threshold energies for its
spontaneous disintegration in various pairs of fragments are shown
in Fig.~\ref{fig.8Be_spectrum} (the data are taken from Ref.~\cite{Tilley}).
As is seen, the $d$-${}^6\mathrm{Li}$ threshold lies $22.2808\,\mathrm{MeV}$
above the ground state (which is also unstable). And this is above the
thresholds for all the other two-body channels, namely,
$t\,{}^5\mathrm{Li}$,
${}^3\mathrm{He}\,{}^5\mathrm{He}$,
$n\,{}^7\mathrm{Be}$,
$p\,{}^7\mathrm{Li}$,
and $\alpha\,\alpha$.\\

When sitting at the nodes of a crystalline lattice, the nuclei oscillate around
their equilibrium positions, but the oscillation energy is negligible on the
nuclear scale. They therefore can be considered as being at rest relative to
each other. In other words, the nuclear pair $d\,{}^6\mathrm{Li}$ in a crystal
is practically at the threshold energy.\\

If $d$ and ${}^6\mathrm{Li}$ in the crystal, overcome huge
Coulomb barrier and fuse, then there is no way back for them. Indeed, for the
resulting excited (resonant) state ${}^8\mathrm{Be}^*$ to decay back into the
channel $d+{}^6\mathrm{Li}$, the deuteron and lithium nuclei must overcome the
same huge Coulomb barrier, while they have practically zero relative kinetic
energy. It is much more easier to decay into one of the channels whose
thresholds are below and where the kinetic energy is above the Coulomb barrier.
For example, in the channel $\alpha+\alpha$ (which lies $0.0918\,\mathrm{MeV}$
below the ground state of the compound nucleus) the relative kinetic energy is
greater than in the $d\,{}^6\mathrm{Li}$ channel by the amount of
$22.2808\,\mathrm{MeV}+0.0918\,\mathrm{MeV}=22.3726\,\mathrm{MeV}$, i.e. is
always well above the barrier.\\

If we restrict our model consideration to only the two-body
channels, then our 8-body problem  involves six-channels, namely,
\begin{equation}
\label{channels}
   d+{}^6\mathrm{Li}\ \rightarrow\ \left\{
   \begin{array}{ll}
   d+{}^6\mathrm{Li}\\[3mm]
   t+{}^5\mathrm{Li}+0.5928\,\mathrm{MeV}\\[3mm]
   {}^3\mathrm{He}+{}^5\mathrm{He}+0.9008\,\mathrm{MeV}\\[3mm]
   n+{}^7\mathrm{Be}+3.3811\,\mathrm{MeV}\\[3mm]
   p+{}^7\mathrm{Li}+5.0257\,\mathrm{MeV}\\[3mm]
   \alpha+\alpha+22.3726\,\mathrm{MeV}
   \end{array}
   \right.
\end{equation}
and all of these channels are open. In other words, if $d$ and
${}^6\mathrm{Li}$ fuse, the final outcome would be one of the pairs on the right
hand side of Eq.~(\ref{channels}) with the energy release shown for each
channel.\\

Around the $(d+{}^6\mathrm{Li})$-threshold, the compound nucleus
${}^8\mathrm{Be}$ has several very wide and overlapping meta-stabele
states (resonances). The
one that is located most closely to the threshold, has the quantum numbers:
(total spin and parity) $J^\pi=2^+$, (isospin) $T=0$, $E=22.2\,\mathrm{MeV}$,
and $\Gamma\approx0.80\,\mathrm{MeV}$ \cite{Tilley}. This wide state is shown
in Fig.~\ref{fig.8Be_spectrum} as a strip of verdical bars.\\

Recent analysis of experimental data, based on the $R$-matrix
parametrization~\cite{Mukhamedzhanov}, showed that at the near-threshold
energies the inelastic collision of $d$ and ${}^6\mathrm{Li}$ mainly leads
to formation of that
resonance ${}^8\mathrm{Be}^*(2^+,0,22.2\,\mathrm{MeV})$. The authors of
Ref.~\cite{Mukhamedzhanov} also showed that this resonance predominantly decays
into the $\alpha\alpha$-channel. They found that the partial
width for such a decay is $\Gamma_\alpha=0.77\,\mathrm{MeV}$, which gives
$\Gamma_\alpha/\Gamma\approx0.96$. In other words, after its formation, this
resonance decays in the $\alpha\alpha$-channel with the probability of 96\%.\\

Ignoring the remaining 4\%, we can assume that, if the fusion of our $d$ and
${}^6\mathrm{Li}$ happens, almost the only outcome is the
$\alpha\alpha$ pair,
\begin{equation}
\label{dLi6to2Alpha}
   d+{}^6\mathrm{Li}\ \longrightarrow\ \alpha+\alpha\ .
\end{equation}
The resulting $\alpha$-particles equally share the energy and momentum.
Therefore the fusion event could be identified by detecting at least one of
the  two $\alpha$-particles moving in the opposite directions with the energies
of $11.1863\,\mathrm{MeV}$.

\subsection{Fusion rate}
\label{sec.fusion_rate}
Nuclear reactions at extremely low energies ($E\sim 10\,\mathrm{keV}$) are
significantly suppressed by the repelling Coulomb forces. The probability $T(E)$
that the colliding nuclei tunnel through the Coulomb barrier is equal to the
ratio $|\psi_E(0)|^2/|\psi_E(R_c)|^2$, where $\psi_E(r)$ is the wave function of
their relative motion at the distance $r$, and $R_c$ is the classical turning
point. It can be shown (see Eq.~(134.10) of Ref.~\cite{Landau}) that for a pure
Coulomb potential, the ratio of $|\psi_E(0)|^2$ to absolute square of the
plane wave is given by
\begin{equation}
\label{psipsiratio}
   T(E)=\frac{2\pi\eta}{\exp(2\pi\eta)-1}
   \quad \mathop{\longrightarrow}\limits_{E\to0}
   \quad 2\pi\eta\exp(-2\pi\eta)\ ,
\end{equation}
where
\begin{equation}
\label{eta}
    \eta=\frac{Z_1Z_2e^2}{\hbar}\sqrt{\frac{\mu}{2E}}
\end{equation}
is the Sommerfeld parameter that involves the nuclear charges, $Z_1$ and $Z_2$,
and the reduced mass $\mu$ of the nuclear pair. For the purpose of estimating,
we can assume that for distances $r>R_c$ the relative motion of the nuclei in
the crystalline cell is described by a plane wave, and thus
Eq.~(\ref{psipsiratio}) gives us the penetration probability for the Coulomb
barrier.\\

In our problem, the nuclei are confined to finite volumes of space in the
crystal. Within its cell, a nucleus moves to and fro, periodically colliding
with the barriers. Each of these collisions is an attempt to tunnel through. If
the size of the cell is $D$ and velocity of the nucleus is $v$, then the
attempts are repeated with the period $2D/v$, i.e. with the frequency
$\nu=v/2D$.\\

Therefore, the number of transitions through the barrier per second (i.e. the
transition rate) is $T\nu$. As we mentioned in Sec.~\ref{sec.FusionChannels}, if
the deuteron (or ${}^6\mathrm{Li}$) manages to pass to the other side of the
barrier, there is no way back (for that, it needs to tunnel once more through
the same barrier, while the penetration probability is very small). This means
that if the penetration takes place, the nuclear system ends up in one of the
inelastic channels given by Eq.~(\ref{channels}). Among these channels, the
reaction (\ref{dLi6to2Alpha}) has the highest probability, $\gamma_\alpha=0.96$.
Therefore, if a deuteron in the crystal oscillates with the energy $E$, the
reaction (\ref{dLi6to2Alpha}) happens with the rate
\begin{equation}
\label{rateW}
   W_d(E)=\frac{T(E)\gamma_\alpha}{2D}\sqrt{\frac{2E}{\mu_d}}\ ,
\end{equation}
where $\mu_d$ is the mass of deuteron. Apparently, the same is valid for the
tunneling of a lithium nucleus, and the corresponding reaction rate
$W_{\mathrm{Li}}(E)$ can be obtained in the same way.

\section{Crystal}

\subsection{Structure}
\begin{figure}[ht!]
\centerline{\epsfig{file=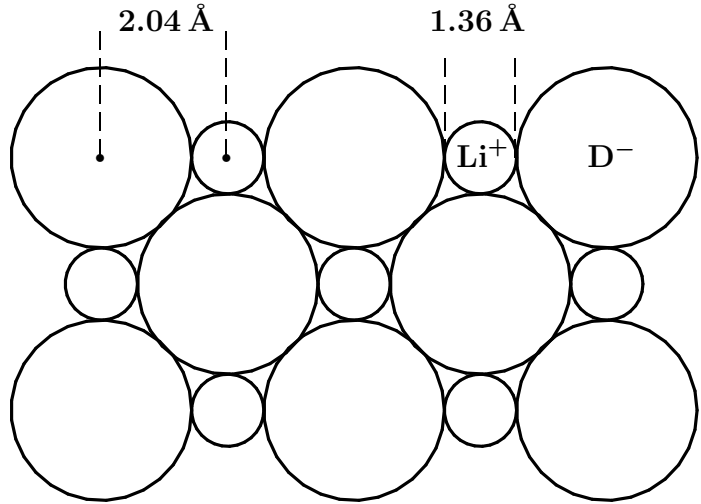}}
\caption{\sf
Schematic picture of one layer of LiD crystal. The indicated distances are
taken from Ref.~\cite{Laplaze}.
}
\label{fig.crystal.plane}
\end{figure}

\begin{figure}[ht!]
\centerline{\epsfig{file=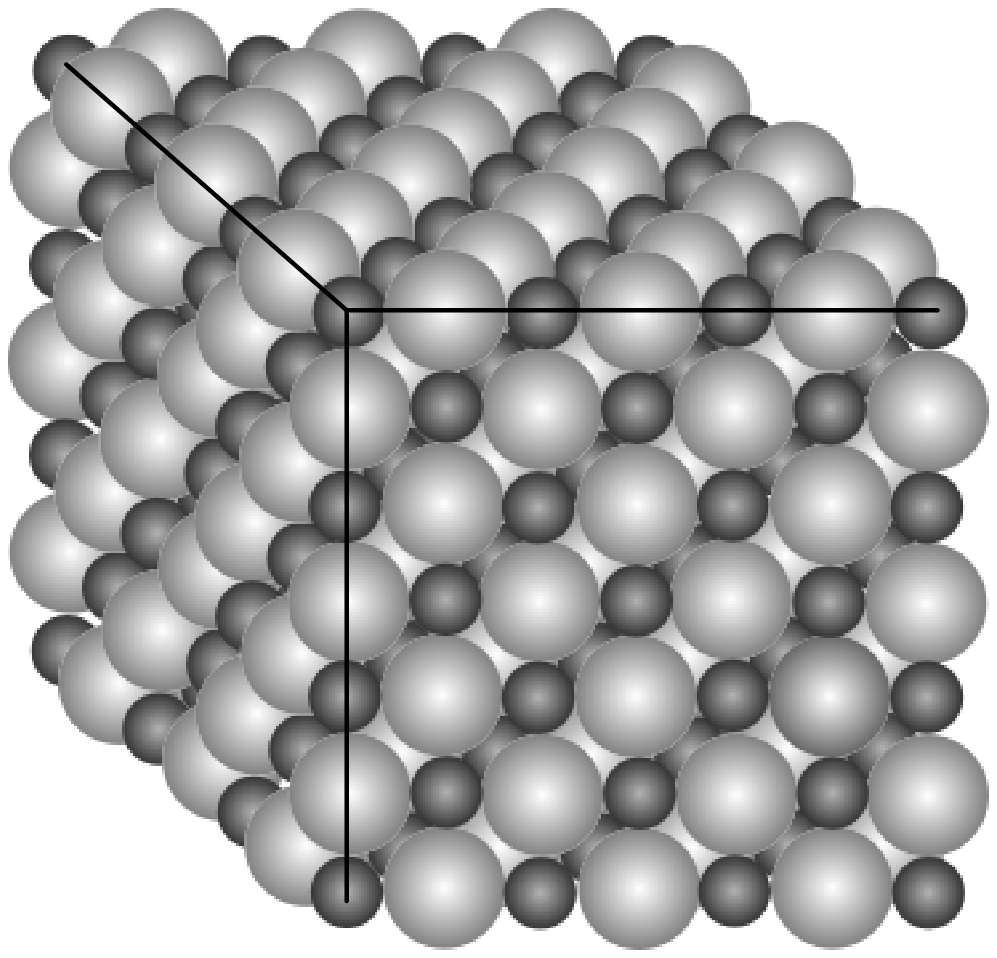}}
\caption{\sf
Schematic 3-dimensional picture of LiD crystal.
}
\label{fig.crystal.3D}
\end{figure}

Lithium hydride is an ionic crystal with simple cubic structure. In each pair
of $\mathrm{Li}$ and $\mathrm{D}$ atoms, one electron is transferred from
the lithium to the deuterium. As a result, the crystal consists of positive
ions of lithium and negative ions of deuterium. The extra electron is loosely
bound to the deuterium, which makes the radius of the ion $\mathrm{D}^-$
approximately twice as much as the radius of $\mathrm{Li}^+$
(see, for example, Ref.~\cite{Laplaze}). Schematically, the structure of lithium
deuteride crystal is depicted in Figures \ref{fig.crystal.plane} and
\ref{fig.crystal.3D}.
\subsection{Inter-nuclear potential}
\label{sec.potential}
\begin{figure}[ht!]
\centerline{\epsfig{file=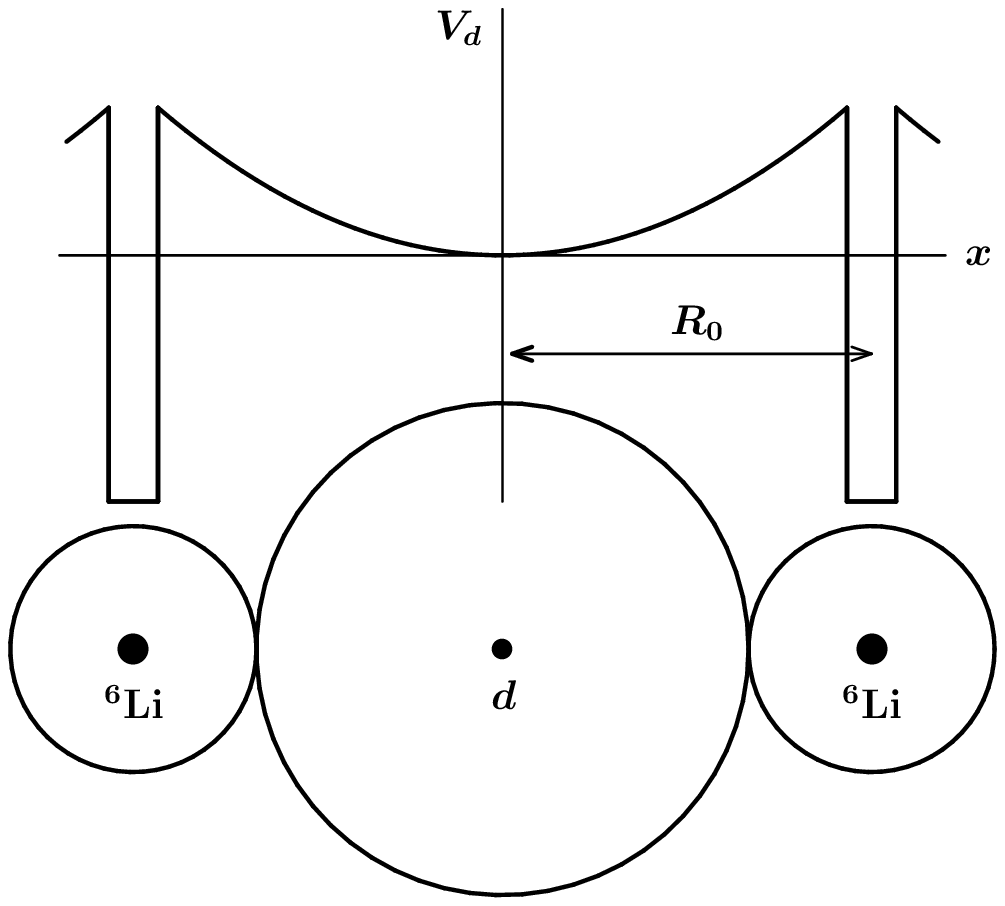}}
\caption{\sf
Schematic picture of the potential energy of a deuteron nucleus in the
Coulomb and nuclear fields of the two neighbouring lithium atoms. The potential
energy is considered along the line connecting the centers of these atoms.
}
\label{fig.crystal.pot}
\end{figure}

If we consider just bare nuclei sitting at the nodes of the lattice, they
repel each other with the Coulomb forces and at short distances attract each
other with the strong forces. The surrounding electrons make the
configuration stable and partly screen the Coulomb repulsion. For a
deuteron nucleus, the neighbouring ${}^6\mathrm{Li}$ nuclei create
the potential profile
schematically shown in Fig.~\ref{fig.crystal.pot}.
There are three orthogonal
axes along which the deuteron moves in such a potential. In order to fuse with
${}^6\mathrm{Li}$, the deuteron have to tunnel through one of the six potential
barriers surrounding it.\\

The height $V_{\mathrm{max}}$ of the barrier can be estimated as the Coulomb
repulsion energy of the charges $Z_1=1$ and $Z_2=3$ at the distance $r_1+r_2$,
where $r_1=2.1424$\,fm~\cite{Mohr} and $r_2=2.5432$\,fm~\cite{Nadjakov} are the
nuclear radii of the deuteron and ${}^6\mathrm{Li}$, respectively. This gives
$$
   V_{\mathrm{max}}\approx
   \frac{Z_1Z_2e^2}{r_1+r_2}\approx0.922\,\mathrm{MeV}\ .
$$
Apparently, the central minimum of the potential shown in
Fig.~\ref{fig.crystal.pot}, is not the same as the zero potential energy for an
isolated
$d\,{}^6\mathrm{Li}$ pair. Indeed, the deuteron is pushed away from both sides.
This effectively lifts the deuteron up against the barrier.\\

If $R_0=2.04\,\mathrm{\AA}$ is the distance between the nuclei $d$ and
${}^6\mathrm{Li}$ in the crystal~\cite{Laplaze}, and $x$ is the shift of the
deuteron from its equilibrium position, then the Coulomb forces acting on it
from the neighbouring lithium nuclei, generate the potential energy:
\begin{equation}
\label{Vx2Coulombs}
    V(x)=\frac{Z_1Z_2e^2}{R_0-x}+\frac{Z_1Z_2e^2}{R_0+x}\ .
\end{equation}
At the equilibrium point ($x=0$) the potential is
\begin{equation}
\label{Vmin}
    V_{\mathrm{min}}=\frac{2Z_1Z_2e^2}{R_0}\approx 42.35\,\mathrm{eV}\ .
\end{equation}
This is how much the deuteron in the crystal is lifted up against the free-space
Coulomb barrier. Actually, the deuteron energy is even a bit higher. This
additional energy, however, is associated with the thermal oscillations and
therefore is small, namely, of the order of $\sim k_BT$, i.e. $\sim
25\,\mathrm{meV}$ at room temperature.\\

The deuteron sitting in the central well of the potential shown in
Fig.~\ref{fig.crystal.pot}, oscillates around its equilibrium point.  In order
to estimate the fusion rate, we need to know the energy levels (spectrum) of its
oscillations, as well as the  distribution
of the statistical ensemble of the deuterons over these levels.
For the estimation purpose, this problem can be simplified if
we approximate the central well by another potential for which both
the spectrum and wave functions are known analytically.\\

Looking at Fig.~\ref{fig.crystal.pot}, one might guess that the best
approximation would be a parabola, i.e. a harmonic oscillator potential. This
however is wrong because the curve shown in that figure is schematic. Actually,
if we accurately plot the function (\ref{Vx2Coulombs}) on the interval
$-R_0+(r_1+r_2)\leqslant x\leqslant R_0-(r_1+r_2)$, it is
practically flat everywhere except for the left and right ends of this
interval, where it quickly raises to $V_{\mathrm{max}}$.
\begin{figure}[ht!]
\centerline{\epsfig{file=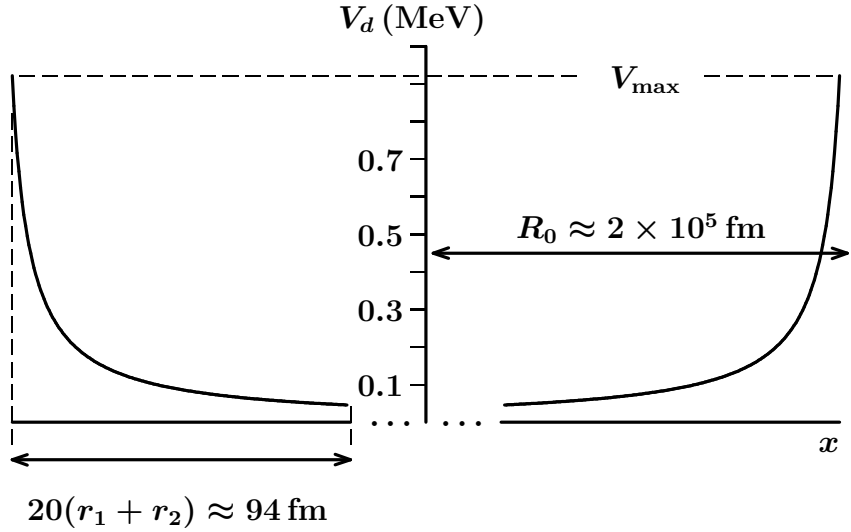}}
\caption{\sf
The potential energy of a deuteron nucleus in the
Coulomb fields of the two neighbouring lithium nuclei. The potential
energy is considered along the line connecting the centers of these nuclei.
There are shown only the leftmost and rightmost segments of this line. The
central part (which is three orders of magnitude longer) is cut out because it
is practically flat.
}
\label{fig.crystal.pot.actual}
\end{figure}
The reason is that $R_0$ is too large as compared to the range of distances
where the Coulomb potential is comparable with $V_{\mathrm{max}}$. Actual
potential-well looks as is shown in Fig.~\ref{fig.crystal.pot.actual}, where we
removed the central part, which is long and almost flat.\\

Of course, in addition to the Coulomb fields generated by the nuclei, there are
also electric fields due to the electron shells of the ions. However,
these fields are weak as compared to the height
$V_{\mathrm{max}}\approx0.922\,\mathrm{MeV}$ of the Coulomb barriers. Indeed,
the binding energy of an electron to the nucleus (and therefore of the
nucleus to the shell) in a hydrogen atom is $\sim13\,\mathrm{eV}$, which is
five orders of magnitude smaller than the height of the barrier. The reason of
relative weakness of the electron field in the crystal is that, in contrast to
the nuclei, the electrons act not as point-like charges but as a space charge of
the electron cloud distributed within a volume of few Angstrom size.\\

For the purpose of estimating, we can assume that the electron cloud is a
uniformly charged sphere.
For a charge $Z_1e$, the interaction potential with the charge $Ze$ uniformly
distributed within a sphere of radius $R$, is (see Eq.~(2-104) of
Ref.~\cite{BohrMottelson})
\begin{equation}
\label{Vsphere}
    V_{\mathrm{sphere}}(r)=\frac{Z_1Ze^2}{2R}\left(3-\frac{r^2}{R^2}\right)\ ,
    \qquad
    r\leqslant R\ ,
\end{equation}
where $r$ is the distance of charge $Z_1e$ from the centre of the sphere. If we
assume (for the sake of estimating) that the electron shell of the ion
$\mathrm{D}^-$ is a uniformly charged sphere ($Z=2$) of the radius
$R_\mathrm{D}=1.36\,\mathrm{\AA}$ ($R_\mathrm{D}=2R_\mathrm{Li}$: see
Fig.~\ref{fig.crystal.plane}), then the
strongest attraction (at $r=0$) for the deuteron from its own electron shell is
$3e^2/R_\mathrm{D}\approx32\,\mathrm{eV}$. In other words, the bottom of the
potential shown in Fig.~~\ref{fig.crystal.pot.actual}, is not exactly flat. At
its centre, it has an additional shallow ``dent'' (potential well) of the depth
$\sim30\,\mathrm{eV}$, which is spherically spread to a distance of
$\sim1.36\,\mathrm{\AA}$.\\
\begin{figure}[ht!]
\centerline{\epsfig{file=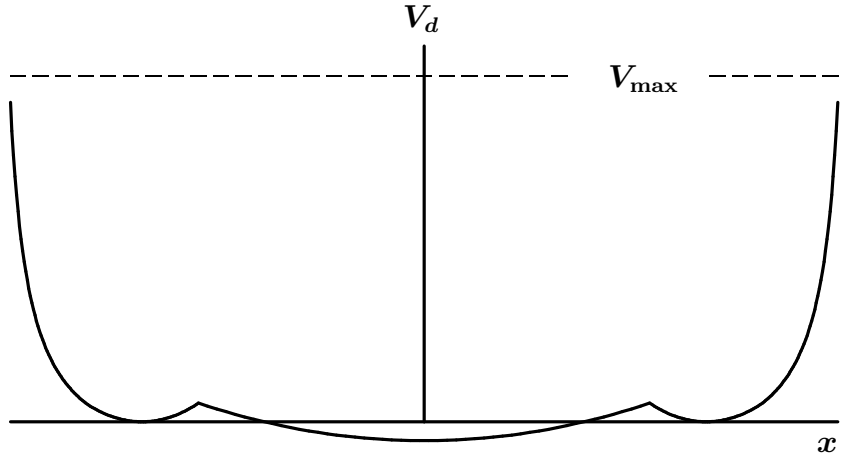}}
\caption{\sf
The potential energy of a deuteron nucleus in the
Coulomb fields of the two neighbouring lithium nuclei modified by the attractive
fields of the electronic shells. This is a schematic picture: the scale is
distorted for the sake of showing the effect of the electron clouds.
As a result of the distortion, this effect is exaggerated here.}
\label{fig.crystal.Vde}
\end{figure}

Similarly, within the electron cloud of the $\mathrm{Li}^+$ ion, there is
another ``dent'' on the potential curve of the depth $\sim60\,\mathrm{eV}$
(because the radius of this cloud is half of the deuterium radius). Therefore
the potential energy of the deuteron nucleus in the crystal (along the line
connecting two neighbouring lithium nuclei) looks like is schematically shown in
Fig.~\ref{fig.crystal.Vde}.\\

It should be noted that despite its depth, the bottom of the lithium ``dent''
is a bit higher relative to
the deuterium one because it is superimposed on the center of the
Coulomb repulsion. It is clear that such a superposition slightly reduces
both the height and the thickness of the Coulomb barrier. Of course the curve
shown in Fig.~\ref{fig.crystal.Vde}, is an exaggeration. Actual depths of the
``dents'' are five orders of magnitude smaller than the height
$V_{\mathrm{max}}$ of the barrier. This means that, if plotted correctly, actual
curve looks practically as a square-well potential of the depth
$V_{\mathrm{max}}$ and width $2R_0$. It should be emphasized that we can
only ignore the electron ``dents'' when consider the motion of the deuteron along
any of the three orthogonal lines, i.e. directly towards one of the
neighbouring lithium nuclei, and when the excitations are
$E\gtrsim 60\,\mathrm{eV}$. In all other directions and at lower oscillation
energies the central ``dent'' is
important because it holds the deuteron in its place.\\

\begin{figure}[ht!]
\centerline{\epsfig{file=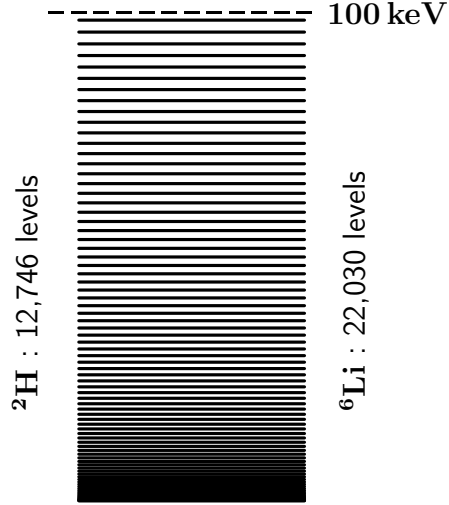}}
\caption{\sf
On the interval $E\in[0,100]\,\mathrm{keV}$, there are 12,746 levels for the
deuteron and 22,030 levels for ${}^6\mathrm{Li}$ in their potential wells.
}
\label{fig.levels}
\end{figure}

Therefore, a good approximation for the potential $V_d(x)$ is a square well.
Moreover, we can replace it with an infinitely deep square well, because the
excitations that we are going to consider ($E\lesssim 100\,\mathrm{keV}$) are
very small as compared to $V_{\mathrm{max}}\sim 1\,\mathrm{MeV}$.
The advantage of using the
infinite square well is that we know both the spectrum and wave functions for
such a potential analytically:
\begin{equation}
\label{sqwell}
   E_n=\frac{\pi^2\hbar^2n^2}{8\mu_d R_0^2}\ ,
   \qquad
   \psi_n(x)=\frac{1}{\sqrt{R_0}}\sin\frac{\pi n(R_0+x)}{2R_0}\ ,
   \qquad
   n=1,2,3,\dots\ ,
\end{equation}
where $\mu_d$ is the deuteron mass. Apparently, everything that was said about
the potential well for the deuteron, is valid for the lithium nucleus as well.
The only parameter that is different in such a case, is the mass of the nucleus.
The ground-state energies for both nuclei in the corresponding potential wells
are very low, namely, $E_1(d)\approx 0.6\,\mathrm{meV}$ and
$E_1(\mathrm{Li})\approx 0.2\,\mathrm{meV}$. The excitation levels are very
dense. On the interval $E\in[0,100]\,\mathrm{keV}$ there are 12,746 and
22,030 levels for the deuteron and ${}^6\mathrm{Li}$, respectively (see
Fig.~\ref{fig.levels}).

\subsection{Crystal under X-rays}
As we have seen, in the lithium deuteride crystal the deuteron is effectively
lifted up against the Coulomb barrier to the energy of $\sim 42\,\mathrm{eV}$.
This gives for the exponential (Gamow) factor
in Eq.~(\ref{psipsiratio}) a negligible value,
$$
    E=42\,\mathrm{eV}\quad\longrightarrow\quad
    \exp(-2\pi\eta)\sim10^{-345}\ .
$$
Such a small value means that the spontaneous fusion in the crystal (although is
possible in principle) is far beyond our capacity to detect it.\\
\begin{figure}[ht!]
\centerline{\epsfig{file=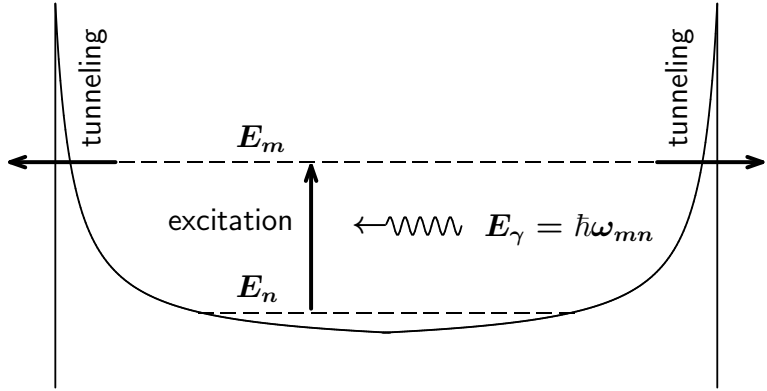}}
\caption{\sf
The deuteron gets the energy from the X-rays and tunnels through the Coulomb
barrier towards the ${}^6\mathrm{Li}$ nucleus.
}
\label{fig.excitation}
\end{figure}

The main idea of the experiment that we describe in the present paper, is to
irradiate the crystal with X-rays, which may excite the oscillations of the
nuclei (near their equilibrium positions) to such a level where they could
tunnel through the Coulomb barrier as
is shematically shown in Fig.~\ref{fig.excitation}. Apparently, the higher the
excitation energy is, the greater the penetration probability would be.

\subsubsection{Stimulated transitions}
Under the influence of an electromagnetic wave, the stimulated excitation
(or de-exitation) rate $p_{mn}$,
i.e. the probability of transition from a state $|\psi_n\rangle$ to another
state $|\psi_m\rangle$ per second, can be found as (taking into account only
the dominant dipole transitions)~\cite{Blokhintsev}:
\begin{equation}
\label{pmn_cos}
   p_{mn}=\frac{4\pi^2Z^2e^2}{\hbar^2}
   \left|\langle\psi_m|x|\psi_n\rangle\right|^2\cos^2\theta
   \rho(\omega_{mn})\ ,
\end{equation}
where $Z=1$ (for deuteron) or $Z=3$ (for lithium), $\theta$ is the angle
between the $x$-axis
and the polarization of the photon, $\omega_{mn}=|E_m-E_n|/\hbar$, and
$\rho(\omega)$ is the volume density of
electromagnetic energy per unit frequency interval. Since we use lithium
deuteride in the form of powder, the angle $\theta$ is random. Relative to the
photon polarization, the $x$-axis can have any orientation uniformly distributed
within full $4\pi$ solid angle. The averaging of $\cos^2\theta$ over
this solid angle gives a factor of $1/3$.\\

The dipole matrix element in Eq.~(\ref{pmn_cos}) can be found using the wave
functions (\ref{sqwell}),
\begin{equation}
\label{mxn_full}
   \langle\psi_m|x|\psi_n\rangle=
   \frac{2R_0}{\pi^2}\left[(-1)^{m+n}-1\right]
   \left[\frac{1}{(m-n)^2}-\frac{1}{(m+n)^2}\right]\ .
\end{equation}
As it should be, this matrix element is non-zero only if $m$ and $n$ have
different parities. In such a case $(-1)^{m+n}-1=-2$ and
\begin{equation}
\label{mxn_parity}
   \langle\psi_m|x|\psi_n\rangle=
   -\frac{16R_0mn}{\pi^2(m^2-n^2)^2}\ .
\end{equation}
Using the first of Eqs.~(\ref{sqwell}), we can replace $m$ and $n$ with the
corresponding energies $E_m$ and $E_n$,
\begin{equation}
\label{mxn_EE}
   \langle\psi_m|x|\psi_n\rangle=
   -\frac{2\hbar^2}{\mu_dR_0}\cdot\frac{\sqrt{E_mE_n}}
   {(E_m-E_n)^2}\ ,
\end{equation}
which gives the following stimulated transition rate (averaged over the photon
polarizations):
\begin{equation}
\label{pmnEmEn}
   p_{mn}=\frac{16\pi^2Z^2e^2\hbar^2}{3\mu_d^2R_0^2}\cdot
   \frac{E_mE_n}{(E_m-E_n)^4}
   \rho(\omega_{mn})\ .
\end{equation}
As is seen, the rate  of high-energy excitations ($E_m\gg E_n$) diminishes
as $\sim E_m^{-3}$.

\subsubsection{Population of energy levels}
The statistical ensemble of deuterons in the crystal is in the thermodynamical
equilibrium, and thus the population $P_n$ of each level (i.e. the probability
that a particular deuteron occupies the level $n$) can be found using the
Boltzmann distribution,
\begin{equation}
\label{BoltzmannPop}
   P_n=
   \frac{\exp\left(-E_n/k_BT\right)}
   {\sum_{j=1}^\infty \exp\left(-E_j/k_BT\right)}\ .
\end{equation}
At the room temperature, $T=300\,{}^\circ\mathrm{K}$, this distribution
gives the following average energy for the oscillations of the deuteron in the
square well potential:
\begin{equation}
\label{Boltzmann}
   \langle E\rangle_d=
   \sum_{n=1}^\infty E_nP_n\approx 14.2\,\mathrm{meV}\ .
\end{equation}
The corresponding energy for the ${}^6\mathrm{Li}$-nucleus is almost the same,
$\langle E\rangle_{\mathrm{Li}}\approx 13.6\,\mathrm{meV}$\,. These small
values are obtained because only a couple of dozens of the lowest levels are
populated
with appreciable probabilities.\\

When the crystal is exposed to X-rays, the statistical ensemble of deuterons (or
similarly ${}^6\mathrm{Li}$ nuclei) is not in the thermodynamical
equilibrium anymore. Its distribution over the energy levels is changing.
The deuterons jump up and down due to absorption and emission of photons.\\

If the flux of external photons is steady, a new dynamical equilibrium is
formed with constant population $P_m(t)=\mathrm{const}$ for each level
$m=1,2,3,\dots$. The time-evolution of the populations can be described by the
so called master equation (see, for example, Ref.\cite{VanKampen}),
\begin{equation}
\label{masterEQ}
    \frac{dP_m}{dt}=\sum_{n\ne m} p_{mn}P_n-\sum_{n\ne m} p_{nm}P_m
    +\sum_{n>m} p_{mn}^{\mathrm{sp}}P_n
    -\sum_{n<m} p_{nm}^{\mathrm{sp}}P_m\ ,
\end{equation}
where $p_{mn}$ is the rate of $(m\gets n)$-transition stimulated by the
external radiation, and $p_{mn}^{\mathrm{sp}}$ is the rate of spontaneous
transition from a higher level $n$ to a lower level $m$. The stimulated
transition rate is given by Eq.~(\ref{pmnEmEn}), and the spontaneous rate (in
the same dipole approximation) can be found as\cite{Blokhintsev}
\begin{equation}
\label{pmnSP}
    p_{mn}^{\mathrm{sp}}=
    \frac{4\omega_{mn}^3Z^2e^2}{3c^3\hbar}
    \left|\langle\psi_m|x|\psi_n\rangle\right|^2\ ,
\end{equation}
involving the same matrix (\ref{mxn_EE}). As we said before, the elements of
this matrix are non-zero only if $m$ and $n$ have different parities, i.e.
$(-1)^{m+n}=-1$.\\

Using Boltzmann distribution (\ref{BoltzmannPop}) as the initial conditions at
$t=0$, we can, in principle, numerically solve the system of differential
equations
(\ref{masterEQ}) up to such time $t$ when all the populations reach constant
values. In practice, however, this is difficult to do. The reason is that the
system (\ref{masterEQ}) consists of too many equations
(see Fig.~\ref{fig.levels}) and therefore becomes
numerically unstable.\\

If we manage to solve the differential equations (\ref{masterEQ}), we obtain
much more information than we actually need. Indeed, such a solution would give
us the full history of how the initial Boltzmann distribution evolves with time.
But we only need the final stationary distribution, when all $P_n(t)$ reach
their constant values and their derivatives on the left hand sides of these
equations become zero. This gives a homogeneous system of
linear equations,
\begin{equation}
\label{homogeneous}
    \sum_nA_{mn}P_n=0\ ,
\end{equation}
with the normalisation condition
\begin{equation}
\label{normalization}
    \sum_nP_n=1\ ,
\end{equation}
where the matrix $A_{mn}$ is composed of the transition rates $p_{mn}$
and $p_{mn}^{\mathrm{sp}}$ as is given in Eq.~(\ref{masterEQ}).\\

It is easy to
proof (using the method of mathematical induction) that the symmetry of the
transition rates with respect to permutations of their subscripts implies that
for each column of matrix $A$, the sum of its elements is zero. These means that
the lines of this matrix are linearly dependent and thus $\det A=0$. In other
words, the linear system (\ref{homogeneous}) always has a unique non-trivial
solution, no matter what the initial distribution is.\\

A homogeneous linear system with the additional condition (\ref{normalization})
and the symmetry properties described above, can be transformed to an equivalent
non-homogeneous system, which can be numerically solved using the Gauss-Seidel
iterative procedure. An algorithm for such a solution can be found in
Ref.~\cite{algorithm}.\\

\begin{figure}[ht!]
\centerline{\epsfig{file=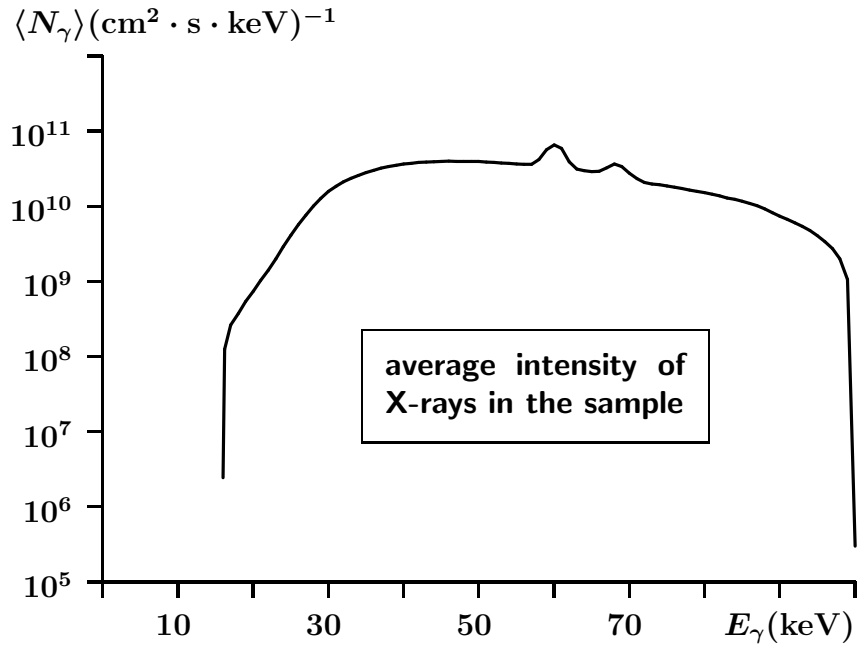}}
\caption{\sf
Number of photons in the unit energy interval, bombarding $1\,\mathrm{cm^2}$
of the target cross section during one second. The number is averaged over
the thickness of the sample.
}
\label{fig.Xspectrum}
\end{figure}
In our experiment, we irradiated the sample with the X-rays whose spectrum
covered the energy interval of $[15,100]\,\mathrm{keV}$. The sample consisted
of many layers of lithium deuteride alternated with polymer detectors
(for details, see Sec.~\ref{sec.sample_preparation}). The
total depth of the sample  along the direction of the radiation was
$13\,\mathrm{cm}$. The radiation intensity attenuated in the sample due
to absorption and the divergence of the rays. In our calculations, we used
the intensity averaged over the depth of the sample. The averaged spectrum
of X-rays is shown in Fig.~\ref{fig.Xspectrum}. It is given in terms of the
number $N_\gamma$ of photons in the unit energy interval falling on the area of
$1\,\mathrm{cm}^2$ per second.\\

Using this spectrum, $N_\gamma(E)$, we calculated the corresponding volume
density $\rho(\omega)$ of electromagnetic energy per unit frequency interval as
follows. In one cubic centimeter, there are those photons that fall on
$1\,\mathrm{cm}^2$ of its side during the time $1\,\mathrm{cm}/c$ needed for the
first bunch of photons to reach the opposite side of the 1\,cm cube. The total
electromagnetic energy in the cube is the number of photons multiplied by
$\hbar\omega$. Dividing this by $\omega$, we obtain the energy density per unit
frequency, $\rho(\omega)=N_\gamma\hbar/c$.\\

\begin{figure}[ht!]
\centerline{\epsfig{file=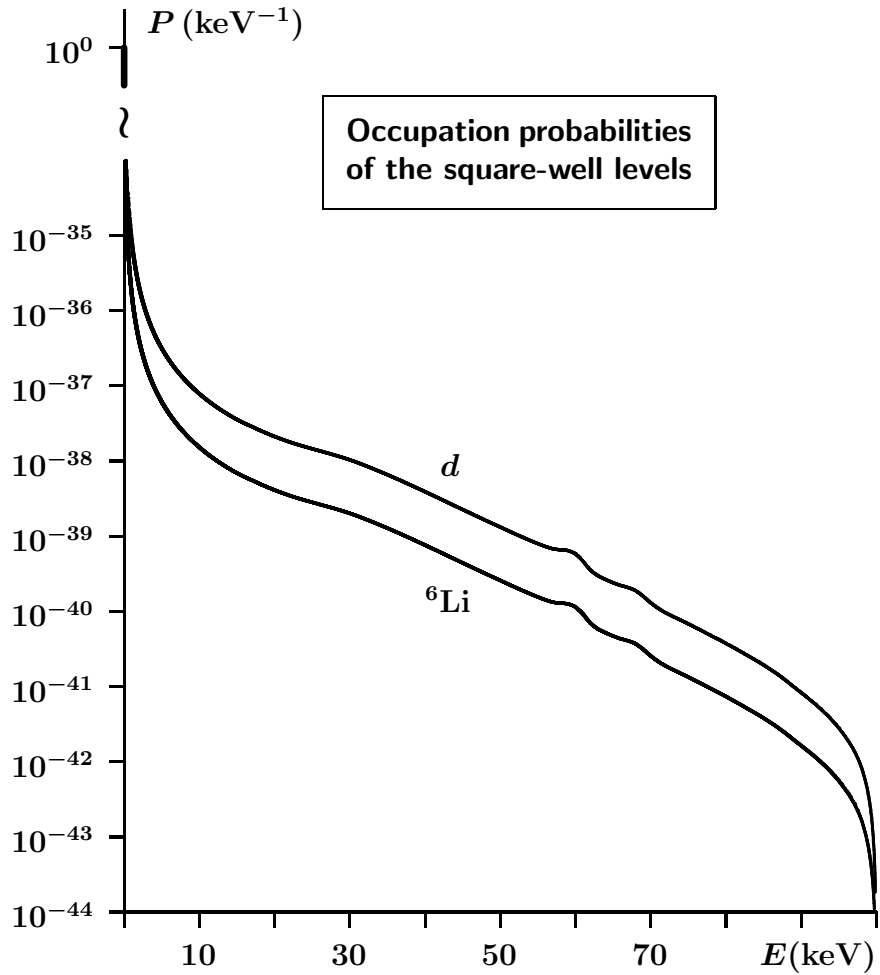}}
\caption{\sf
Probabilities (per unit energy interval) of populating the levels in
the square-well potential for the deuteron and ${}^6\mathrm{Li}$ nuclei,
when the crystal is exposed to the X-rays with the spectrum shown in
Fig.~\ref{fig.Xspectrum}.
}
\label{fig.population}
\end{figure}
This density is needed in finding the stimulated transition rates
(\ref{pmnEmEn}). With these rates and the rates
(\ref{pmnSP}) of the spontaneous transitions, we solved the linear system
(\ref{homogeneous}) and thus found the stationary probability distribution
for occupying various energy levels in the potential square-well. This was done
for both the deuteron and ${}^6\mathrm{Li}$ nuclei (the only difference is
the mass). The resulting distributions are shown in Fig.~\ref{fig.population}.

\subsection{Fusion induced by X-rays in the crystal}
\label{sec.fusion_induced_X}
When the crystal is exposed to the X-rays, the nuclei oscillating within the
square-well potentials, acquire  some kinetic energy.
As a result, the probability for them to penetrate
through the Coulomb barrier increases. The corresponding fusion rate for a
deuteron (and similarly for lithium) is given by Eq.~(\ref{rateW}).\\

In that equation, the energy $E$ does not have a definite value, but can be any
$E_n$ of the square-well spectrum for the deuteron (or lithium, which is
different), with the probability distributions $P_d(E)$ and $P_{\mathrm{Li}}(E)$
shown in Fig.~\ref{fig.population}. This is similar to the reactions in stellar
plasma \cite{Rolfs}, where velocities of the colliding nuclei are distributed
according to Maxwell's probability density. Therefore the observable fusion rate
is the following average
\begin{equation}
\label{w_av}
   \langle W_i\rangle=\sum_nW_i(E_n^{(i)})P_i(E_n^{(i)})\ ,
\end{equation}
where $i$ stands for either $d$ or $\mathrm{Li}$.\\

Each deuteron has six neighbouring lithium nuclei and similarly each lithium
nucleus has six hydrogen isotopes surrounding it. However, the crystal is not
pure $\mathrm{D}{}^6\mathrm{Li}$ compound. In some nodes of the lattice it can
be a proton (instead of $d$) or ${}^7\mathrm{Li}$ (instead of
${}^6\mathrm{Li}$). In our experiment, it was used a sample containing
$M=0.61\,\mathrm{g}$ of lithium hydride powder with natural isotope composition
of lithium (mass fractions of ${}^6\mathrm{Li}$ and ${}^7\mathrm{Li}$ being
$f_6=0.0759$ and $f_7=0.9241$, respectively) and enriched with the hydrogen
isotope ${}^2\mathrm{H}$ ($f_1=0.02$ and $f_2=0.98$).\\

Therefore, if you find a hydrogen atom in the crystal, it is a deuterium with
the probability of $f_2$. And a lithium atom has in its centre the
${}^6\mathrm{Li}$ isotope with the probability of $f_6$. Therefore, a deuteron
can find around itself a ${}^6\mathrm{Li}$ nucleus with the probability of
$6f_6$, and for a ${}^6\mathrm{Li}$ isotope the probability to find a deuteron
nearby is $6f_2$.\\

Since there are two possibilities for the same fusion event to happen: either
the deuteron  or the lithium gets through the barrier, the total (``bulk'')
fusion rate for the whole crystal can be found as a sum of the corresponding
contributions:
\begin{equation}
\label{w_bulk_corrected}
     R=6f_6N_d\langle W_d\rangle+
     6f_2N_{\mathrm{Li}}\langle W_{\mathrm{Li}}\rangle\ .
\end{equation}
where $N_d$ and $N_{\mathrm{Li}}$ are the numbers of available deuterons and
${}^6\mathrm{Li}$ nuclei.\\

The effective molar mass $m$ of our crystal-powder is
\begin{equation}
\label{molar_mass}
    m=f_11\,\mathrm{g}+f_22\,\mathrm{g}+f_66\,\mathrm{g}+
      f_77\,\mathrm{g}\ .
\end{equation}
The number of hydrogen and lithium atoms (any isotopes) is the same, namely,
$(M/m)N_A$, where $N_A$ is the Avogadro number. This gives
\begin{equation}
\label{d_and_Li6_numbers}
  N_d=\frac{M}{m}N_Af_2\ ,
  \qquad
  N_{\mathrm{Li}}=\frac{M}{m}N_Af_6\ ,
\end{equation}
and therefore
\begin{equation}
\label{w_bulk_final}
     R=6\frac{M}{m}N_Af_2f_6\Bigl(\langle W_d\rangle+
     \langle W_{\mathrm{Li}}\rangle\Bigr)\ .
\end{equation}
Numerical calculations with the X-ray spectrum shown in
Fig.~\ref{fig.Xspectrum},
give the following results for the single-nucleus rates:
\begin{equation}
\label{2body_rates}
   \langle W_d\rangle\approx 2.4\times10^{-26}\,\mathrm{s}^{-1}\ ,
   \qquad
   \langle W_{\mathrm{Li}}\rangle\approx
   4.6\times10^{-27}\,\mathrm{s}^{-1}\ .
\end{equation}
The corresponding ``bulk'' rate for the whole sample is
\begin{equation}
\label{bulk_rate_number}
   R\approx 5.2\times10^{-4}\,\mathrm{s}^{-1}\ .
\end{equation}
For the total exposure time in our experiment, $t=111.466$ hours, we therefore
should have expected to register $N\sim207$ fusion events.

\section{Experiment}
As it follows from the theoretical estimate (see previous Section), the fusion
events we are looking for, are rare (roughly, one event every half an hour).
This fact determines the sample design and the choice of the way of registering
the events. The low rate of the reaction implies that we need to expose the
sample to the X-rays for a relatively long time (at least a hundred hours) to
get a statistically meaningful result. The detector should be able to
efficiently distinguish the fusion events from all possible background signals
and to accumulate the information during whole exposure period.\\

Based on this, we used the polymer track detectors CR-39 in direct contact with
the active material (lithium hydride). Each fusion event results in a pair of
$\alpha$-particles moving in the opposite directions with the same energy of
$11.1863\,\mathrm{MeV}$. These particles leave tracks in the polymer material,
which can be identified after the experiment is completed. Since only few
hundred events are expected, the probability that different tracks overlap each
other is practically zero. The background tracks left by other charged
particles, can be easily excluded using specific properties of the tracks
belonging to the $\alpha$-particles from the fusion events (see
Sec.~\ref{sec.detectors}).
\subsection{Sample preparation}
\label{sec.sample_preparation}
Lithium hydride ($\mathrm{LiH}$) is a solid crystalline compound, usually
available in the form of chunks or powder. It is extremely hygroscopic,
absorbing water from the air, via the chemical reaction
$$
     \mathrm{LiH + H_2O\ \longrightarrow\ LiOH + H_2}\ .
$$
For our experiment, this is a destructive process  and therefore we avoided
the contacts of $\mathrm{LiH}$ with atmospheric air by all available means.  All
the manipulations with the crystals were done in an anaerobic chamber
(glove-box) filled with dry $\mathrm{CO_2}$ gas. Besides that, some desiccants
were present in the chamber.\\

\begin{figure}[ht!]
\centerline{\epsfig{file=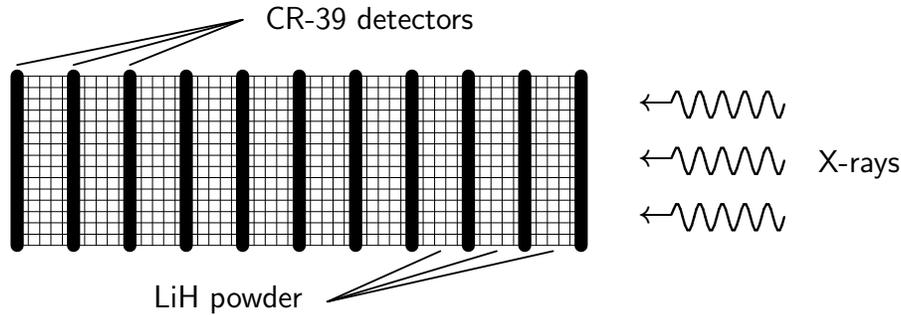}}
\caption{\sf
Schematic picture of a sample unit composed of alternating layers of
lithium-deuteride powder and the plastic detectors (CR-39). Whole assembly
includes several units like this, placed  parallel to each other inside a
hermetic PVC container. The distance between the detector plates is 1\,cm. The
total number of the detectors in whole assembly is 85. Each detector plate is a
square of the area 1cm$\times$1cm with 1\,mm thickness. The X-rays are directed
perpendicular to the plates.
}
\label{fig.slojka}
\end{figure}
The crystalline chunks were ground into powder form.
This powder was placed between square plates ($1\,\mathrm{cm^2}$ each) of the
plastic detectors (CR-39), as is schematically shown in Fig.~\ref{fig.slojka}.
We put in parallel several units like the one shown in this Figure, and the
whole assembly was enclosed in a hermetic PVC container. The total number of the
detector plates was 85.\\

The lithium hydride we used, was not pure $\mathrm{D{}^6Li}$-compound. The
isotope composition of lithium atoms was practically natural, namely, 92.41\% of
$\mathrm{{}^7Li}$ and 7.59\% of $\mathrm{{}^6Li}$. As far as the hydrogen atoms
are concerned, the substance was enriched with deuterium up to 98\%.

\subsection{X-ray source and spectrum}
To generate the X-rays, we used a XYLON Y.TU 225-D02 tube. The tube was operated
with a potential difference of 100\,kV and the current of 30\,mA. In order to
find out what was the spectrum of the rays, we used available data for the
spectrum of this source, measured at the same voltage, but with a much lower
current, namely, $0.1\,\mu\mathrm{A}$, and at a different distance. We simply
scaled these data to actual current and distance, used in our experiment.\\

The length of our sample was 13.2\,cm. Passing through the sample, the X-rays
attenuate both because of interaction with the material and because of spreading
(the rays are not collinear). As a result the intensity of the radiation is not
uniform through the sample. In our theoretical estimate, we used average
intensity of the X-rays in the sample. The corresponding spectrum (obtained by
scaling the original data and averaging them) is shown in
Fig.~\ref{fig.Xspectrum}.

\subsection{Background}
We identify the fusion events by registering the $\alpha$-particles that emerge
from the decay of the resulting ${}^8\mathrm{Be^*(22.2\,MeV)}$ resonance.
However, there are other (background) sources of $\alpha$-particles that might
be distorting our counting. Among them the most common is radon (and its decay
products) that is present practically everywhere. In addition to that, our
measurements were done at the premises of NECSA (Nuclear Energy Corporation
South Africa) near Pretoria, where a nuclear reactor was operating. Although an
appropriate radiation shielding was used, we cannot exclude that some thermal
neutrons from the reactor could reach our sample and generate $\alpha$-particles
in collisions with ${}^6\mathrm{Li}$ via the reaction
\begin{equation}
\label{neutron}
    n+ {}^6\mathrm{Li}\ \longrightarrow\ \alpha + t + 4.78\,\mathrm{MeV}\ .
\end{equation}

Our detectors were kept isolated from the atmospheric air all the time. And
since
$\alpha$-particles cannot penetrate through a plastic container, we can safely
ignore the possibility of counting false events due to radon and its progeny.
The only source of possible false events is the reaction (\ref{neutron}).
However, the $\alpha$-particles from the fusion events have a much higher energy
(11.1\,MeV) than those  from this reaction. This fact provides us with a
reliable way of distinguishing the fusion and false events. When analysing the
detectors, we only counted the tracks left by particles with the energy higher
than 6\,MeV (for the details see Sec.~\ref{sec.detectors}).

\subsection{Detector processing}
\label{sec.detectors}
The fusion events were registered by identifying the appropriate tracks left by
$\alpha$-particles in the solid state nuclear track detectors CR-39. Original
tracks are very narrow channels where the structure of the plastic material is
damaged by the particle. These channels become much wider and visible under a
microscope magnification, after etching. For etching, we used a 6.25 mol/$\ell$
solution of NaOH at 70 ${}^\circ$C.\\

This solution removes the plastic material not only from the channels, but from
all the surfaces. However the rate of etching at the damages is higher. There
are certain empirical formulae that enable one to calculate the etching rates at
various conditions (see Ref.~\cite{Azooz} and the references therein). It is
therefore possible to numerically model the shapes of the tracks for different
etching times. For such a modelling, we used the standard code TRACK\_TEST
developed by the authors of Ref.~\cite{Nikezic}.\\

\begin{figure}[ht!]
\centerline{\epsfig{file=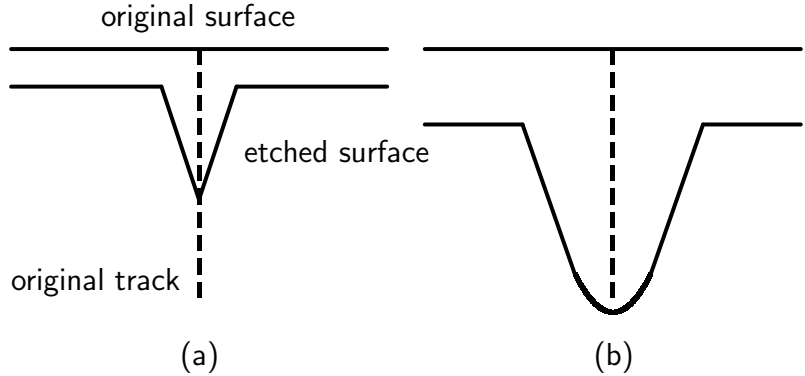}}
\caption{\sf
Schematic representation of the etching process for two different durations: (a)
Etching time is not sufficient to reach the bottom of the original track; (b)
Rounding of the bottom of the pit after the track end has been reached.
}
\label{fig.tracks}
\end{figure}
Depending on its kinetic energy, the $\alpha$-particle penetrates into the
plastic material and damages it to a certain depth (typically a dozen of
microns). When etched, such a track forms a pit in the shape of a sharp cone, as
is shown in Fig.~\ref{fig.tracks}(a). The longer we etch it, the deeper and
wider this cone becomes. This goes on until the depth of the pit becomes equal
to the penetration length. After that the pit only becomes wider and retains
practically the same depth. As a result, the bottom of the pit transforms from a
sharp cone to a smooth spherically concave surface, as is schematically shown in
Fig.~\ref{fig.tracks}(b).\\

The sharp cone and concave shapes have different optical properties. A smooth
concave surface reflects the light and even focus it to a small dot, which is
not possible for a sharp cone. Therefore just looking into the etched tracks, we
can distinguish ``finished'' (track end has been reached) and ``unfinished''
tracks. Using the TRACK\_TEST software, we can calculate the length of the track
for any given energy $E_\alpha$ of the $\alpha$-particle as well as calculate
the depth of the pit for a given etching time $t$. Therefore we can choose $t$
such that all the tracks for $E_\alpha$ higher than certain threshold
$E_{\mathrm{min}}$, form sharp cone pits, while for $E_\alpha<E_{\mathrm{min}}$
the pits reflect the light.\\

Since we need to exclude the tracks resulted from the reaction
(\ref{neutron}), where the maximal energy is 4.78\,MeV, we could choose the
threshold energy to be 5\,MeV. However, to be absolutely sure, we used
$E_{\mathrm{min}}=6\,\mathrm{MeV}$. This corresponds to etching time
$t=8$\,hours. After the etching, we looked for completely dark pits (sharp
cones) with certain  diameter of the opening. These tracks could not be done by
anything else but the $\alpha$-particles from the fusion reaction.

\subsection{Efficiency of the detecting}
\label{sec.efficiency}
It should be noted that we cannot register all the fusion events that occur in
the sample. First of all, the $\alpha$-particles from the reactions that take
place far away from a detector plate, cannot reach the detector because the
distance they can go through in any solid media is very short.
Moreover, even if they reach a
detector surface from afar, they loose a significant part of the kinetic energy
and thus cannot be registered if the remaining energy is less than
$E_{\mathrm{min}}=6\,\mathrm{MeV}$.\\

Our calculations show that in the lithium-hydride the original energy of
11.1\,MeV of the $\alpha$-particle is reduced to the minimally acceptable energy
of 6\,MeV on the distance of $x_{\mathrm{max}}=0.095$\,mm. This means that, when
we count the events, we can only register the events happening within a thin
layer of the lithium-hydride of 0.095\,mm thickness, which is in contact with
the detector.
Since in total we have 85 detectors and for each of them the surface area in
contact with LiH is 0.95$\times$0.95\,cm${}^2$ (most of them have LiH on both
sides), the total volume of LiH, from which we register the $\alpha$-particles,
is 0.77\,cm${}^3$ and the corresponding mass is 0.61\,g. This effective mass of
the active material was used in our theoretical estimate of the fusion rate.\\

This is not the end of the story. The efficiency of the detection is reduced
further by the fact that the $\alpha$-particles are emitted isotropically in all
directions. For example, if the fusion event happens at the maximal acceptable
distance from the detector and the $\alpha$-particle moves at a non-zero angle
relative to the normal to the detector surface, the actual distance it has to
pass in LiH is greater than the $x_{\mathrm{max}}$ and thus it looses too much
energy. Similar losses take place at any distance within the thin layer, when
the angle is too large. Moreover, the tracks with the angle to the normal
greater that $45^\circ$ cannot be identified using our simple method.\\

Estimating all possible losses, we conclude that we can only register of about
40\% of the events happening in 0.61\,g of the material. This means that from
the number of detected events we can only derive a lower bound for the reaction
rate. This is sufficient for a first and simple experiment. Actually, the main
goal of our experiment was to establish if the fusion in the crystal could be
induced by X-rays or not.

\section{Results and discussion}
The sample composed of alternating layers of lithium-hydride and the plastic
detector-plates (as is described in Sec.~\ref{sec.sample_preparation}), was
exposed to the X-ray radiation for the duration of 111.466 hours. Then, after
the etching of the detectors, we identified (in total on all 85 plates) 88
tracks that belonged to the $\alpha$-particles from the fusion reaction
(\ref{dLi6to2Alpha}). Since we could not register all the fusion events (some
of them were missing as is explained in Sec.~\ref{sec.efficiency}), actual
number $N$ of the events (happened in 0.61\,g of LiH) is greater, $N>88$. Taking
into account that the substance we used was not pure lithium-deuteride (see
Sec.~\ref{sec.fusion_induced_X} for explanations), we conclude that under the
used electromagnetic radiation (the spectrum is given in
Fig.~\ref{fig.Xspectrum}) the fusion rate for a single
$d-{}^6\mathrm{Li}$ pair is
$$
      R_{d{}^6\mathrm{Li}} > 1.19\times10^{-26}\,\mathrm{s}^{-1}\ .
$$
This result is of the same order of magnitude as was predicted by our
theoretical estimate (\ref{2body_rates}). The total number of events $N=207$,
theoretically expected for our experimental conditions and duration, is in
accordance with the actual observation, $N>88$.\\

Apparently, the fusion rate turned out to be too low for any possible
applications of this process in the energy production. However, the
electromagnetically induced nuclear reactions in crystals can be used in a
different way. In principle, this offers a new way of measuring the cross
sections (or the astrophysical S-factors) of such reactions (not only fusion) at
extremely low energies, which are not accessible in the direct collision
experiments.\\

Of course, such an approach would require a more rigorous theoretical
description of all the processes involved. In particular, the electron screening
has to be taken into account in a proper way. This could be done by
appropriately modifying the barrier penetration factor and using an accurate
model (instead of the square well) for the potential schematically shown in
Fig.~\ref{fig.crystal.Vde}.
There is no doubt that taking into account the electron clouds is not an easy
task. Indeed, the configuration of their space charge is dynamically changing
when the nuclei move. It is therefore needed some theoretical effort before the
nuclear reactions in crystals could be used for measuring the corresponding
cross sections at low energies.\\

At present, the only alternative to the direct collision experiments at low
energies is the so called ``Trojan Horse'' method, where the projectile is
carried to the close proximity of the target as a cluster inside another nucleus
(the ``horse''). This ``horse'' may have the kinetic energy well above the
Coulomb barrier and thus neither its penetrability nor the screening effects
play any role. When near the target, the projectile is detached from the
``horse'' and interacts with the target at a low energy while practically all
the original kinetic energy is taken away by the ``empty horse'' (see
\cite{Spitaleri}, and the references therein).\\

Although in the ``Trojan Horse'' method the complicated screening effects are
avoided, it still relies on an involved theoretical analysis. Therefore, an
additional experimental method for determining the S-factors would be of help.
The ``Trojan Horse'' and the crystal methods would complement each other.\\

Experimentally, the main advantage of using crystals is the fact that here the
effective flux of the nuclei is extremely dense. Indeed, roughly the Avogadro
number of projectiles collide with the Avogadro number of target nuclei. With
such a density (as we have seen in our experiment) even the reactions with very
low probabilities can be observed. In order to narrow the interval of the
collision energies, a synchrotron radiation can be used instead of the X-rays.


%
%

\end{document}